\documentclass[fleqn,usenatbib]{mnras}

\usepackage{newtxtext,newtxmath}

\usepackage[T1]{fontenc}
\usepackage{ae,aecompl}

\usepackage{graphicx}	
\usepackage{amsmath}	
\usepackage{amssymb}	
\usepackage{etoolbox}
\makeatletter
\patchcmd\@combinedblfloats{\box\@outputbox}{\unvbox\@outputbox}{}{%
   \errmessage{\noexpand\@combinedblfloats could not be patched}%
}%
 \makeatother

\newcommand{\Msun}{\,{\rm M}_{\sun}}

\newcommand{\Gyr}{\,{\rm Gyr}}

\newcommand{\FeH}{\lbrack{\rm Fe/H}\rbrack}
\newcommand{\A}{\lbrack{\rm \alpha /Fe} \rbrack}
\newcommand{\Oxy}{\lbrack{\rm O/Fe} \rbrack}

\title[Alpha abundances]{The $\A$-$\FeH$ relation in the E-MOSAICS simulations: its connection to the birth place of globular clusters and the fraction of globular cluster field stars in the bulge}

\author[M. E. Hughes et al ]{Meghan E. Hughes$^{1,2}$\thanks{E-mail: M.Hughes1@2013.ljmu.ac.uk},
Joel L. Pfeffer$^{1}$,
Marie Martig$^{1}$,
Marta Reina-Campos$^{3}$, 
\newauthor
Nate Bastian$^{1}$, 
Robert A. Crain$^{1}$,
and J. M. Diederik Kruijssen$^{3}$
\\
$^{1}$Astrophysics Research Institute, Liverpool John Moores University, 146 Brownlow Hill, Liverpool L3 5RF, UK\\
$^{2}$European Southern Observatory, Karl-Schwarzschild-Stra{\ss}e 2, 85748, Garching, Germany \\
$^{3}$Astronomisches Rechen-Institut, Zentrum f\"{u}r Astronomie der Universit\"{a}t Heidelberg, M\"{o}nchhofstra{\ss}e 12-14, 69120 Heidelberg, Germany\\
}

\date{Accepted XXX. Received YYY; in original form ZZZ}

\pubyear{2019}

\begin{document}
\label{firstpage}
\pagerange{\pageref{firstpage}--\pageref{lastpage}}
\maketitle

\begin{abstract}
The $\alpha$-element abundances of the globular cluster (GC) and field star populations of galaxies encode information about the formation of each of these components. We use the E-MOSAICS cosmological simulations of $\sim L\star$ galaxies and their GCs to investigate the $\A$-$\FeH$ distribution of field stars and GCs in 25 Milky Way-mass galaxies.
The $\A$-$\FeH$ distribution of GCs largely follows that of the field stars and can also therefore be used as tracers of the $\A$-$\FeH$ evolution of the galaxy.
Due to the difference in their star formation histories, GCs associated with stellar streams (i.e. which have recently been accreted) have systematically lower $\A$ at fixed $\FeH$. Therefore, if a GC is observed to have low $\A$ for its $\FeH$ there is an increased possibility that this GC was accreted recently alongside a dwarf galaxy.
There is a wide range of shapes for the field star $\A$-$\FeH$ distribution, with a notable subset of galaxies exhibiting bimodal distributions, in which the high $\A$ sequence is mostly comprised of stars in the bulge, a high fraction of which are from disrupted GCs. We calculate the contribution of disrupted GCs to the bulge component of the 25 simulated galaxies and find values between 0.3-14 per cent, where this fraction correlates with the galaxy's formation time. The upper range of these fractions is compatible with observationally-inferred measurements for the Milky Way, suggesting that in this respect the Milky Way is not typical of $L^{*}$ galaxies, having experienced a phase of unusually rapid growth at early times.

\end{abstract}



\section{Introduction\label{1}}
The element abundances of stars and globular clusters (GCs) are powerful tools with which to extract information about the time and place of their formation, giving us an insight into how galaxies form and assemble. The element abundances of Milky Way (MW) GCs are often used to assess whether they formed in the MW or in a satellite galaxy that was later accreted. A powerful set of abundances are those of the $\alpha$ elements. The abundance ratio of $\alpha$ elements to iron, $\A$, is an important tracer of the relative contributions of Type II/Ia supernovae (SN), since only Type II SN contribute to the production of ${\rm \alpha}$ elements whereas both contribute to iron \citep{Wheeler1989}. This makes $\A$, together with $\FeH$, a good tracer of the enrichment history of a galaxy. For example, a star or GC with a high $\A$ at fixed $\FeH$ indicates that its progenitor gas was enriched primarily with ${\rm \alpha}$ elements synthesised and promptly released by Type II supernovae, whilst incorporating relatively little iron synthesised by Type Ia supernovae \citep{Wheeler1989,McWilliam1997}. In many galaxies, low $\FeH$ stars that formed before Type Ia SN enriched the interstellar medium (ISM) show a relatively constant $\A$. There is then a `knee' in the $\A$-$\FeH$ distribution, where stars begin to form from the Type Ia SN enriched material, followed by a downwards trend of decreasing $\A$ as the ISM continues to be enriched by Type Ia SN. Lower mass galaxies do not self enrich as fast as higher mass galaxies and therefore the position of the `knee' is shifted to lower $\FeH$ (e.g. \citealt{Pritzl2005,Tolstoy2009}).\par  

The $\A$-$\FeH$ distribution of field stars in cosmological simulations has been addressed in several recent studies. \citet{Mackereth2018} used the EAGLE simulations to investigate the $\A$-$\FeH$ distribution around the solar neighbourhood of 133 MW-like galaxies in terms of their stellar mass and kinematics. They found a diversity in the shape of the distributions, noting that only five per cent of them show a bimodal $\A$ distribution, similar to that exhibited by the MW. The simulations indicate that this bimodality, in particular the appearance of a high-$\alpha$ sequence, occurs in galaxies that experience rapid growth at early epochs in response to a period of vigorous star formation triggered by the atypically early formation of their dark matter halo. The low-$\alpha$ sequence is then formed by a subsequent prolonged period of less intense star formation. The authors therefore concluded that the MW also underwent a rapid early growth, making it an atypical $L^*$ galaxy. \citet{Grand2018} found $\A$ bimodality in the disc populations of six MW-sized halos in the Auriga simulations. Consistent with \citet{Mackereth2018}, they attribute bimodality in the inner disc to a central starburst (caused by a gas rich merger), followed by less intense star formation. In the outer disc, they further attribute $\A$ bimodality to early $\alpha$-rich star formation in a gas disc, followed by a shrinking of the disc that lowers the star formation rate. It is of particular interest that both studies attribute a high-$\alpha$ sequence to an early, rapid star formation episode \citep[also see][]{kruijssen19b}.\par 

In the MW GCs exhibit similar $\A$ to field stars at fixed $\FeH$ (e.g. \citealt{Pritzl2005}). However, there are some Galactic GCs, such as Ruprecht 106 (Rup 106) and Palomar 12 (Pal 12) that have relatively low $\A$ ratios for their $\FeH$ values with respect to both the MW field stars and other GCs. It has been hypothesised that these GCs have been captured from dwarf galaxies with a different chemical enrichment history to the MW \citep{Lin1992,Sneden2004,Pritzl2005,Forbes2010}. A similar offset is also seen when comparing stars in dwarf galaxies with the MW field stars \citep{Pritzl2005,Tolstoy2009}.\par

In the field stars, element abundances can be useful for finding groups of stars that were born in the same molecular cloud: this is called `chemical tagging', a concept introduced by \citet{Freeman2002}. Finding the stars that once belonged to bound clusters has become a major topic with the recent advances in Galactic surveys such as APOGEE \citep{Majewski2017}, Gaia-ESO \citep{Gilmore2012}, RAVE \citep{Steinmetz2006,Zwitter2008,Siebert2011} and GALAH \citep{DeSilva2015,Buder2018}. With these surveys it is possible to tag chemically hundreds of thousands of stars, making it possible to identify stars likely to have once been members of the same star cluster (e.g. \citealt{PriceJones2019}). This same technique may be used to identify stars that once belonged to the same dwarf galaxy, providing an insight into the accretion history of the MW. On a larger scale, chemical tagging to find disrupted GCs and dwarf galaxies gives some clues about the early star formation process of the galaxy and its dynamical history \citep{Ting2015}. \par

In addition to identifying stars that were born in the same molecular cloud, it is also interesting to consider more broadly the fraction of field stars that originated in GCs. If the fraction of stars formed in bound star clusters varies with the surface density of star formation \citep[e.g.][]{Kruijssen2012}, then the disrupted GC contribution to the thin disk, thick disc and bulge offers clues as to how each of these components formed. Stars that have formed within GCs can be identified by exploiting star-to-star abundance variations within GCs (e.g. \citealt{Fernandez2019}), known as multiple populations (e.g. \citealt{Gratton2004,Bastian2018}). This has been carried out in the halo of the MW, where 2-3 per cent of halo stars were found to exhibit chemical signatures seen in GC stars \citep{Martell2011,Carollo2013,Martell2016,reinacampos19b}. These studies then attribute 4-17 per cent of halo stars as once being part of a GC, depending on the GC formation mechanism and the fraction of enriched stars initially within GCs. \cite{Schaivon2017} carried out a similar analysis in the Milky Way bulge in a specific metallicity range of $\FeH < -1$ and, by finding nitrogen-enriched stars, concluded that 14 percent of the stellar mass of the bulge came from disrupted GCs.\par 

In a companion paper, we investigate the GC contribution to the halo of the 25 simulated MW-mass galaxies from the E-MOSAICS simulations. We find a median of 0.3 per cent of the mass in halo field stars formed in GCs, indicating that the disruption of GCs plays a sub-dominant roll in the build-up of galaxy stellar haloes \citep{reinacampos19b}.\par

In this paper we continue the study of $\alpha$ abundances in cosmological simulations using the 25 zoom-in E-MOSAICS simulations described by \citet{Pfeffer2018} and \citet{kruijssen19} (for which the relevant details are discussed in Section \ref{2}), which enable us to follow the formation, evolution and disruption of GCs alongside the evolution of their host galaxy. We discuss the differences and similarities between the field stars and GCs and the in-situ and ex-situ GCs in Section \ref{3}.  In Section \ref{4}, we investigate the formation and disruption of GCs in the $\A$-$\FeH$ plane. In Section \ref{5}, we present how the amount of GC disruption (particularly that in the bulge) can be related to the shape of the field star $\A$-$\FeH$ distribution and subsequently the formation time of the galaxy. Finally, in Section \ref{6} we present the conclusions of this work.

\section{Simulations \label{2}}
For this work we use the E-MOSAICS (MOdelling Star cluster population Assembly in Cosmological Simulations within EAGLE) suite of 25 zoom-in simulations of MW mass galaxies \citep{Pfeffer2018,kruijssen19}. E-MOSAICS couples the MOSAICS subgrid model for star cluster formation and evolution \citep{Kruijssen2011,Pfeffer2018} to the EAGLE (Evolution and Assembly of GaLaxies and their Environments) galaxy formation model \citep{Schaye2015,Crain2015}.These simulations follow the co-formation and evolution of galaxies and their star cluster populations in a cosmological context.

We focus here on the specific aspects of the simulations relevant to this work, namely the chemical abundances, the cluster formation efficiency (CFE) and cluster mass loss. We refer the interested reader to \citet{Schaye2015},  \citet{Pfeffer2018} and \citet{kruijssen19} for full details of the EAGLE and MOSAICS models, respectively. \par
To follow the formation of a galaxy halo, the SUBFIND algorithm \citep{Springel2001,Dolag2009} is used to identify subhaloes (galaxies) in the simulations, from which galaxy merger trees were constructed \citep[see][for details]{Pfeffer2018}. The merger trees enable us to assign a parent galaxy (we define `parent galaxy' as the subhalo the particle was bound to prior to forming a star) to each GC and we can therefore label a GC's formation as `in-situ' (its parent galaxy is the main progenitor) or `ex-situ' (its parent galaxy merged with the main progenitor). \par
EAGLE tracks the abundances of the 11 elements most important for radiative cooling. Following \citet{Segers2016} and \citet{Mackereth2018}, we use $\Oxy$ as a proxy for $\A$, since oxygen dominates the mass budget of $\alpha$ elements. Abundance ratios $[\mathrm{a}/\mathrm{b}]$ are calculated as
\begin{equation}\label{eq:1}
\mathrm{[a/b]} = \log_{10}\Bigg( \frac{X^{\mathrm{a}}}{X^{\mathrm{b}}}\Bigg) -  \log_{10}\Bigg( \frac{X^{\mathrm{a}}_{\odot}}{X^{\mathrm{b}}_{\odot}}\Bigg).
\end{equation}
We adopt solar fractions of $X^{\mathrm{O}}_{\odot} / X^{\mathrm{Fe}}_{\odot} = 4.98$ and $X^{\mathrm{Fe}}_{\odot} /X^{\mathrm{H}}_{\odot} = 0.0016$ from \citet{Wiersma2009}, consistent with the cooling tables used in EAGLE.\par

MOSAICS adopts a star cluster formation model based on observations of young star clusters, under the assumption that young star clusters, open clusters and GCs share a common formation mechanism (see reviews by \citealt{Longmore2014,kruijssen2014,Bastian2016}).  When a star particle forms in the simulations, a fraction of its mass may form a star cluster population. Once this mass has been assigned to cluster formation, cluster masses are stochastically drawn from a Schechter \citep{Schechter1976} inital cluster mass function (ICMF) with an environmentally dependent truncation mass. \footnote{ So as to not impose an upper limit on the cluster mass, the cluster masses are allowed to exceed the mass of the particle. This is justified because occasionally, the truncation mass of the ICMF exceeds the mass of a stellar particle in these high resolution simulations.}.

Cluster populations for each star particle form with a local, environmentally-varying cluster formation efficiency (CFE; i.e. the fraction of stars formed in bound clusters, \citealt{Bastian2008}) and initial cluster mass function according to the models of \cite{Kruijssen2012} and \cite{Reina-Campos2017}, respectively. This cluster population inherits the age and chemical composition of its parent stellar particle. \par

The clusters lose mass due to stellar evolutionary mass loss (in common with field stars) and dynamical processes. Mass loss through stellar evolution is tracked for each stellar particle by the EAGLE model \citep{Wiersma2009} and the lost mass is donated to neighbouring gas particles. For dynamical evolution, mass loss from tidal shocks and two-body relaxation is included, as described in detail by \citet{Kruijssen2011}, this redistributes the mass internally within the stellar particle from the bound to the field populations. Clusters are evolved down to 100 $\Msun$, after which they are assumed to be completely disrupted. Disruption of clusters by dynamical friction is applied to clusters at every snapshot in post-processing (since stellar particles may host clusters of different masses, see \citealt{Pfeffer2018} for details). In this Section and Section \ref{3} we define a surviving GC as a star cluster in the simulations with a present day mass greater than $10^{5} \Msun$ and an age greater than 2 Gyr. In Section \ref{4} our discussion is focused on the \textit{initial} GC population that has now disrupted so we define a GC as a star cluster with an initial mass greater than $10^5 \Msun$ and an age greater than 2 Gyr.
We use this age cut since we want to look for disrupted GCs in the field by utilising multiple populations, a phenomenon which is known to exist in GCs older than 2 Gyr \citep{Martocchia2018}.\par
The E-MOSAICS simulations reproduce many key observables of the MW and M31 GC populations, such as the specific frequency, the number of low metallicity GCs ($\FeH < -1$) and the radial distribution of GCs. We direct the reader to fig. 2 and the accompanying discussion of \citet{kruijssen19} for an overview. In addition, the high mass end of the simulated globular cluster mass function (GCMF) is in good agreement with that of the MW and M31. However, the simulations produce too many low mass clusters (particularly at $ < \mathrm{10^5} \Msun$), most likely due to under disruption \citep{Pfeffer2018}. In this work we account for this by limiting the GCs to $\mathrm{> 10^5} \Msun$ and providing extreme cases of complete disruption where necessary. Another observable reproduced by the E-MOSAICS simulations is the `blue tilt', where more massive GCs are typically more metal-rich. Using the E-MOSAICS simulations \citet{Usher2018} reinterpreted the `blue tilt' as an absence of massive metal-poor GCs as a consequence of less-massive and therefore less-metal-rich galaxies not reaching sufficiently high interstellar gas surface densities to form massive GCs (see \citealt{kruijssen19} for a similar but subtly different interpretation).
The E-MOSAICS simulations also provide predictions for future observations, \citet{Reina-Campos2019} find a median lookback time of GC formation to be $10.7 \Gyr$, approximately $2.5 \Gyr$ earlier than that of the field stars. As a result, proto-GC formation is predicted to be the most prevalent between $\mathrm{2 < z < 3}$ (consistent with the peak of star formation history \citealt{Madau2014}), a prediction which can be tested with JWST.

\section{The $\alpha$ abundances of globular clusters and field stars\label{3} }
\begin{figure*}
	\includegraphics[width=\linewidth]{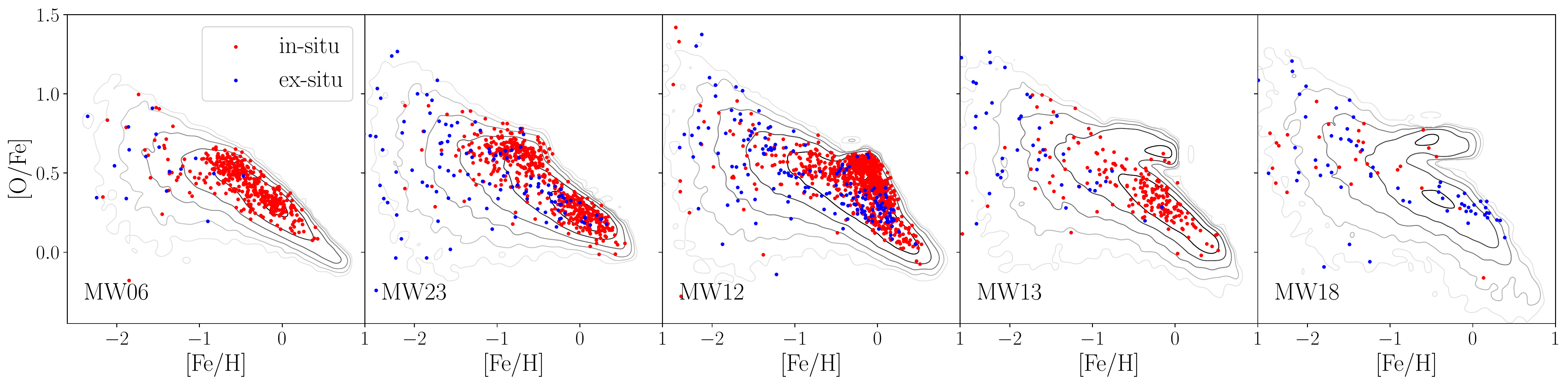}
    \caption{Five of the E-MOSAICS galaxies in $\A$-$\FeH$ space, chosen to illustrate the diversity of the $z=0$ $\A$-$\FeH$ distributions. The contours represent the field stars and the points represent the in-situ (red) and ex-situ (blue) GCs.  }
    \label{fig:row.pdf}
\end{figure*}

The $\alpha$ element abundances of GCs have been used to establish whether a GC is likely to have been formed in-situ or ex-situ, under the assumption that a GC with low $\A$ at a fixed $\FeH$ (relative to the MW's field stars) indicates an ex-situ origin (e.g. \citealt{Pritzl2005}). The motivation for this follows from the assumption that GCs formed ex-situ did so in a satellite galaxy with a longer gas consumption timescale than the main progenitor. In particular, low mass dwarf galaxies are expected to transition to low $\A$ at lower $\FeH$ than more massive galaxies \citep{Matteucci1990,Tolstoy2009}. The underlying assumption to this classification is that the GC formation history is broadly representative of the field star formation history in all galaxies.\par

\cite{Pritzl2005} used a compilation of Galactic GCs with high-fidelity stellar abundance measurements and compared their $\A$ abundances with those of the field stars. They find that  GCs follow the abundances of field stars reasonably well, with a few exceptions. In particular, they suggest that Ter 7,  Pal 12 and Rup 106 have an extra-galactic origin based on their lower $\A$ abundances. This is also suggested in other studies where Ter 7 \citep{DaCosta1995} and Pal 12 \citep{Dinescu2000} are inferred to be associated with the tidally disrupting Sagittarius dwarf galaxy. It has been suggested that Rup 106 is of extragalactic origin although its parent galaxy is still debated \citep{Bellazzini2003,Law2010,Forbes2010,Massari2019}. Other MW GCs with low $\A$ suggesting extragalactic origin include NGC 5694 \citep{Lee2006,Mucciarelli2013} and Pal 1 \citep{Monaco2011,Sakari2011}. There is also evidence that the GCs in the Fornax dwarf galaxy have lower $\A$ when compared to the MW GCs \citep{Larsen2012}. Also, \citet{Cohen2004} and \citet{Tautvaisiene2004} concluded that the known Sagittarius GCs follow the $\A$ trend of the known Sagittarius field stars.\par

We present in Fig. \ref{fig:row.pdf} a subset of the E-MOSAICS galaxies with a range of shapes in $\A$ - $\FeH$ to highlight key points in the differences and similarities between galaxies. These are MW06, MW12, MW13, MW18 and MW23 in Table 1 of \citet{kruijssen19}. The contours represent the field stars, and the GCs are overplotted as red or blue points depending if they formed in-situ or ex-situ respectively. Unless otherwise stated, when we refer to `field stars' in this work we are referring to all stellar particles that are bound to the main subhalo. When calculating the mass of the field stars, the mass of the globular cluster population associated with the stellar particle is omitted. In all the galaxies, both the field stars and the GCs show a decline of $\A$ with increasing $\FeH$.   From left to right, the panels show field star distributions ranging from a smooth decline to being clearly bimodal at fixed metallicity (in the range -1<[Fe/H]<0).  \cite{Mackereth2018} used the EAGLE simulations to investigate the $\A$ abundances of the field stars around the solar radius (thus excluding stars in the bulge) in a sample of MW-like galaxies. They also found bimodality in 5 per cent of their galaxies and attribute the appearance of bimodality to a phase of rapid growth early in the galaxy's formation history - we investigate this explanation and its relation to GCs in Section \ref{5}. \par

\begin{figure*}
	\includegraphics[width=\linewidth]{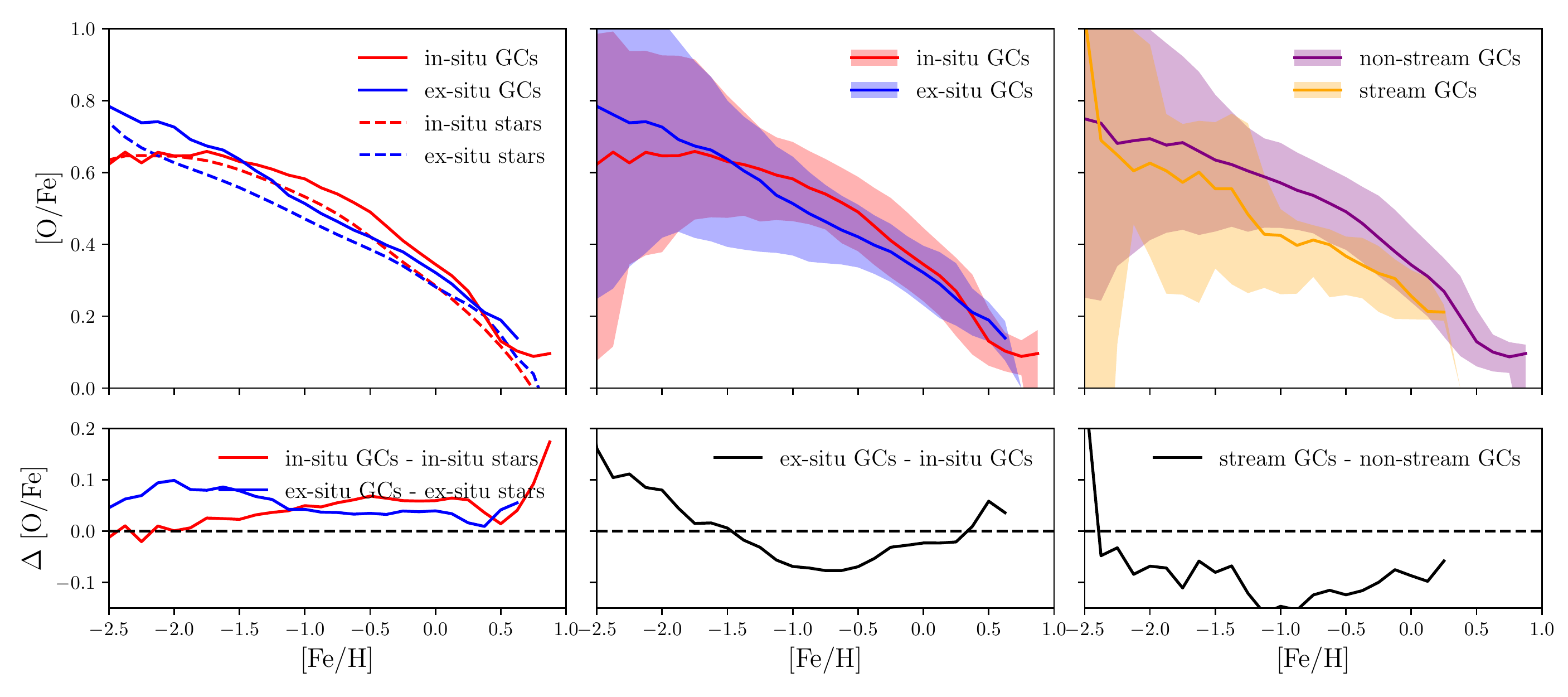}
    \caption{Top left: the median $\Oxy$ as a function of metallicity of the in-situ and ex-situ field stars and GCs in all 25 galaxies. Top middle: the median $\Oxy$ as a function of metallicity of the in-situ and ex-situ GCs in all 25 galaxies. Top right: the median $\Oxy$ as a function of metallicity of the GCs on streams and not on streams in the 15 MW-like galaxies identified in \citet{Hughes2019}.  The shaded regions show the 16th and 84th percentiles and the lines are a running median of $\Oxy$ in $\FeH$ bins of 0.5 dex with a difference of 0.125 dex between each bin. The bottom panels show the differences in $\Oxy$ between the subsets of stars or GCs for each $\FeH$ bin. All panels only include bins with more than 5 GCs to avoid poor sampling issues.}
    \label{fig:insitu_exsitu.pdf}
\end{figure*}

The first thing we note is that the abundances of the GCs closely trace the abundances of the field star population; however, similarly to the MW there are some clear exceptions. All of the galaxies shown in Fig. \ref{fig:row.pdf} host a small population of GCs that have low $\A$ for a given $\FeH$. However, contrary to what is assumed for the MW, this population of GCs is not universally ex-situ. In E-MOSAICS GCs follow the abundance trends of the field stars by construction (there cannot be a GC without a star particle), but where the GCs form and how they evolve in the simulation may impart biases on the properties of the star particles that still hold a GC at $z=0$. We investigate these points further by stacking all 25 MW-like galaxies and looking for systematic trends in Fig. \ref{fig:insitu_exsitu.pdf}. \par

To test whether GCs follow the abundance trends of the field stars in our simulations, we show the median $\Oxy$ for fixed bins of $\FeH$ of the in-situ (red) and ex-situ (blue) GCs and field stars of the 25 galaxies (top left panel of Fig. \ref{fig:insitu_exsitu.pdf}). Both field stars (dashed lines) and GCs (solid lines) show a decrease in the median $\Oxy$ with increasing $\FeH$, a similar trend to that seen in the MW field stars and GCs \citep[e.g.][]{Hayden2015}. We also find that the GCs follow the general trend of the field stars, but they are offset to higher $\Oxy$. We quantify this difference in the bottom left panel, where we show the difference in median $\Oxy$ between stars and GCs for both the in-situ and ex-situ populations. This panel shows that the GC $\Oxy$ is always greater than that of the field stars by $\sim0.05$ dex. We expect GCs to show higher $\Oxy$ because, in the MOSAICS model, GCs are formed in high density environments, and high density environments induce shorter gas consumption times (Appendix \ref{A}, \citealt{tacconi18},\citealt{Mackereth2018}, fig. 4), which in turn leads to higher $\Oxy$.
 The gas consumption time is an estimate of the time a star-forming gas particle resides in the ISM before becoming a star particle. It can vary significantly across a single galaxy due to variations in pressure. A correlation between gas consumption time and $\A$ is not necessary, because such correlation only arises if a parcel of gas is self-enriched (i.e., there is no dilution from gas infall, the Fe and $\alpha$ elements produced by stellar evolution are not ejected to large distances, and there is no large-scale radial mixing of gas within the galaxy). In the EAGLE simulation, a correlation between gas consumption time and $\A$ has been demonstrated by \cite{Mackereth2018} (their fig. 4).  This arises because metals produced by stellar evolution are returned locally (using the SPH kernel), and because gas consumption timescales are similar to the timescales  of the Type IaSNe delay time distribution. \par

To further study the difference in the $\alpha$ abundances of the in-situ and ex-situ GCs, we show their median $\Oxy$ values as a function of $\FeH$ in the middle panel of Fig. \ref{fig:insitu_exsitu.pdf}, where the shaded region shows the 16th and 84th percentile range. Although the ex-situ GCs show, on average, systematically lower $\Oxy$,  the distributions heavily overlap and the difference between the medians is smaller than the $1\sigma$ ranges. The middle bottom panel of Fig. \ref{fig:insitu_exsitu.pdf} shows the difference between the in-situ and ex-situ GC median $\Oxy$. There is a range of $\FeH$ ($-1.2 < \FeH < -0.25$) in which ex-situ GCs have a lower median $\Oxy$ than in-situ GCs, but outside of this range $\Oxy$ is similar or ex-situ GCs have higher $\alpha$ enhancement (particularly for $\FeH < -1.8$).
Therefore we cannot say definitively that ex-situ GCs show lower $\alpha$ abundances at all $\FeH$. However, the ex-situ GCs in these simulations are identified as any GC which has been accreted onto a central galaxy over its full formation history. This means that some of the ex-situ GCs were formed in progenitors that were accreted very early on in the galaxy's formation history, and consequently they would most likely be identified as in-situ GCs in chemical and kinematic studies. Dwarf galaxies accreted early tend to have more rapid formation histories (in terms of the time it takes them to reach a maximum mass) than those accreted late \citep{Mistani2016}, therefore we would expect them to have higher $\A$ abundances. Therefore, it would be prudent for us to examine an alternative definition of an ex-situ GC to facilitate a more direct comparison with observations of GCs in the Milky Way. \par

The most direct evidence for accretion in the Milky Way comes in the form of stellar streams, therefore we complements the in-situ/ex-situ comparison with a stream/non-stream comparison. For that, we use the sample of stellar streams in 15 of the E-MOSAICS MW-mass galaxies from \cite{Hughes2019}, which were visually identified in 2D projections of the stellar particles of accreted galaxies. We show their $\Oxy$ abundances in the right panel of Fig. \ref{fig:insitu_exsitu.pdf} and quantify the differences in the bottom right panel. This panel shows that the stream GCs have consistently lower $\Oxy$ abundances than the non-stream GCs. At $\FeH = -1$ the difference between the stream and the non-stream GCs is double that of the difference between the ex-situ and the in-situ GCs. Therefore we can conclude that if we observe a GC in the halo of a galaxy that belongs to a stream, there is a high probability that it will be $\alpha$-poor relative to the main GC population. This supports the conclusion of works that state that ex-situ GCs should exhibit lower $\A$ at fixed $\FeH$ than in-situ GCs (e.g. \citealt{Pritzl2005}), however we would revise this conclusion to state that we can distinguish \textit{recently} accreted GCs this way, i.e. that high $\A$ does not necessarily imply that a GC formed in-situ, as at low metallicity most GCs form with high $\A$. We can also conclude that if we were to find a low $\A$ GC in the halo of a galaxy it is likely to have been accreted relatively recently and could be a signpost for the presence of an associated disrupting dwarf galaxy. The lower $\alpha$ abundance of these accreted stars and GCs is not driven by their ex-situ origin in itself, but by the fact that they formed and accreted recently.

\section{The $\A$ - $\FeH$ distribution of field stars and its connection to the formation and disruption of globular clusters \label{4}}

\subsection{Cluster formation and disruption across the $\A$-$\FeH$ plane}

\begin{figure*}
	\includegraphics[width=\linewidth]{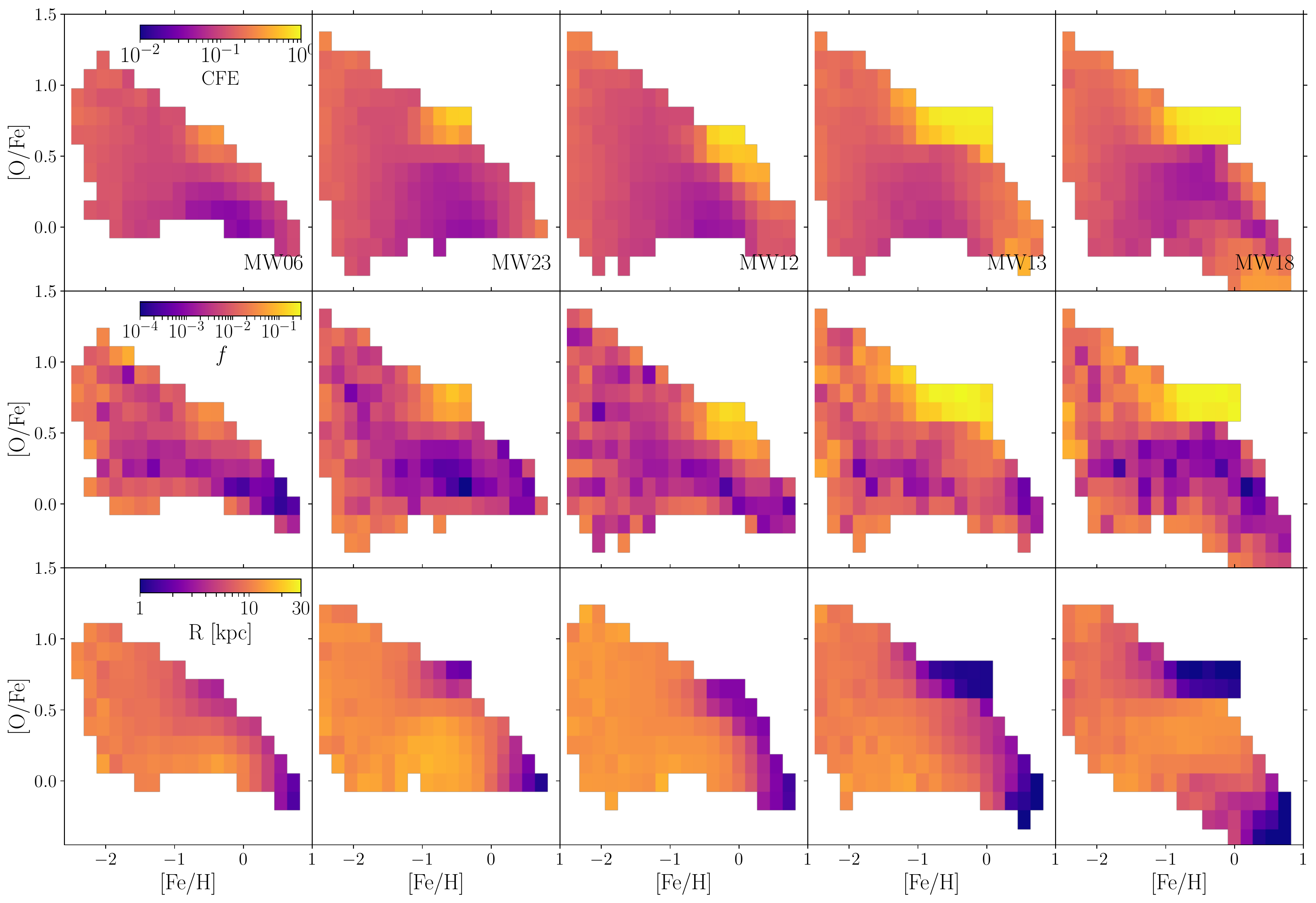}
    \caption{The $\Oxy$ - $\FeH$ relation for five of the simulated galaxies. Each panel shows a 2D histogram of the stars in the galaxy coloured by,  the cluster formation efficiency (CFE), the fraction of stars which were born in GCs but now reside in the field ($\mathit{f}$) and the galactocentric radius of the stars at $z =0$ (R). In the last row, only the stars from the inner 30 kpc are shown.}
    \label{fig:big_plot.pdf}
\end{figure*}

Section \ref{3} shows us that the $\alpha$ abundances of a galaxy's GCs encode information about the formation and assembly of the galaxy's GC population. We now focus on what we may learn about the contribution of disrupted GCs to the galaxy's field star population from their $\alpha$ abundances. In Fig. \ref{fig:row.pdf} MW13 and MW18 show a bimodal  $\Oxy$ distribution in their field stars at $\FeH \approx -0.25$, however, there are relatively few GCs in the high $\A$ sequence compared to the low $\A$ sequence. This motivates investigation of the $\A$-$\FeH$ plane in terms of GC formation and disruption.\par 

\subsubsection{GC formation}
We first focus on the formation of GCs through the CFE. The CFE governs the fraction of star formation that yields bound clusters (see \citealt{Adamo2018} for a recent review). The CFE increases with star formation rate surface density \citep{Adamo2015} and in the E-MOSAICS simulations it scales with the natal gas pressure. E-MOSAICS uses the environmentally dependent description of the CFE from the \citet{Kruijssen2012} model, which relates the fraction of star formation into bound stellar clusters to the properties of the interstellar medium (ISM) - bound clusters form most efficiently at the high density end of the hierarchically structured ISM. \par

The first row of Fig. \ref{fig:big_plot.pdf} shows a 2D histogram of the stellar particles in the $\Oxy$-$\FeH$ space coloured by the CFE associated with their birth cloud (or `natal gas'). The CFE ranges from a few percent to 80 percent, depending on the location in $\A$-$\FeH$ space.
As the stellar $\A$-$\FeH$ distribution becomes more bimodal (Fig. \ref{fig:row.pdf}, from left to right), the high CFE feature in the galaxy (Fig. \ref{fig:big_plot.pdf}, second row) becomes more pronounced in the high-$\alpha$ sequence. The CFE is dependent on the density of the natal gas of the stellar particle (through the natal pressure, due to the equation of state imposed on dense, star forming gas), with higher densities leading to a higher CFE. Therefore the high-$\alpha$ sequence must form from material with increased natal gas pressure, meaning that the stars form from gas with short gas consumption times (see Appendix \ref{A} and \citealt{Mackereth2018} for details).

However, as discussed earlier, the bimodal galaxies MW13 and MW18 have a lack of GCs in their high $\A$ sequence, even though this is the same area of $\A$ - $\FeH$ space where the CFE is the highest. This means that although there is a clearly defined area in this plane where the galaxy is forming a large number of GCs, these GCs do not survive to $z=0$. Therefore we now investigate GC disruption across the $\A$-$\FeH$ plane.

\subsubsection{GC disruption}
The region of high CFE in the $\A$-$\FeH$ plane tells us that the stars and GCs which form in this region do so in a high density environment. However, tidal shocks are also more prevalent in high density environments and can efficiently disrupt the nascent cluster. This is the `cruel cradle effect' described by \cite{Kruijssen2012a} and means that where a galaxy is likely to form many clusters, it is also likely to disrupt them. Dynamical friction also removes many of the most massive clusters that are not disrupted by tidal shocks, particularly in the centres of galaxies. The combination of the cruel cradle effect and dynamical friction explains the absence of GCs in the same location in $\A$-$\FeH$ space where the CFE is high in Fig. \ref{fig:big_plot.pdf}. \par

We can make a measurement of the fraction of the stars that formed in bound massive clusters similar to present day GCs but now reside in the field. For this purpose we revise slightly our definition of a GC to that of a star cluster with an \textit{initial mass} greater than $ 10^{5} \Msun$ and an age greater than 2 Gyr. The second row of Fig. \ref{fig:big_plot.pdf} shows a 2D histogram of all the stellar particles in the galaxy weighted by the fraction of their mass that once belonged to a GC that has since dissolved into the field star population, 
\begin{equation}\label{eq:2}
f = \dfrac{\Sigma_i^{N_*} \left(M_{\rm GC, init}\times {\rm SML} - M_{\rm GC, final}\right)}{\Sigma_i^{N_*} M_{*, field}},
\end{equation}
where $M_\mathrm{GC, init}$ is the initial total mass in GCs, $M_\mathrm{GC, final}$ is the final total mass in GCs, $M_\mathrm{*, \ field}$ is the final total mass field star population in the stellar particles and the factor $\mathrm{SML}=M_{*, \mathrm{final}}/M_{*, \mathrm{init}}$ corrects the initial total mass in GCs for stellar evolutionary mass loss (such that we are only considering dynamical mass loss). We include mass loss from tidal shocks and two-body relaxation but not the complete removal of clusters by dynamical friction. Dynamical friction is omitted since we assume that this mass will quickly sink to the centre of the galaxy potentially contributing to the nuclear star cluster\footnote{Any potential increase in tidal disruption due to a shrinking orbit cannot be captured in the present model.} (e.g. \citealt{Antonini2013}). Therefore these stars would not be easily identifiable through chemical tagging studies of the field star population of the Galaxy.\par

Some of the galaxies show a clearly defined region of $\Oxy$-$\FeH$ space where up to 30 per cent of the field stars were born in GCs. This region of high $f$ is the same as the region of high CFE and overlaps with the high $\Oxy$ sequence in the galaxies which show bimodality in the top row. Therefore, we show that in some galaxies, some stars are born in high density regions of star forming gas, which means that for a given $\FeH$ their $\Oxy$ will be high. Due to the high densities it is likely that these stars will form in bound clusters, but because of the `cruel cradle effect' a high fraction of these clusters will also be fully or partially disrupted. Hence, in some galaxies, there is expected to be a region in $\A$-$\FeH$ space where a high fraction of field stars were originally born in GC-like clusters. This has implications for chemical tagging studies and will be discussed in detail in Section \ref{5}.\par

\subsection{Galactocentric position in the $\A$-$\FeH$ plane}
We investigate whether there is a radial dependence on where stellar particles will be distributed in $\A$-$\FeH$ space. In the bottom row of Fig. \ref{fig:big_plot.pdf} we show a 2D histogram of the stellar particles within 30kpc of the centre of the galaxy, coloured by their galactocentric (spherical) radius. The region of $\Oxy$-$\FeH$ space that shows the highest fraction of disrupted GC stars ($f$) resides at the centre, or bulge, of the galaxy. This is expected since the centres of galaxies usually show the highest pressures due to the radial pressure gradient of the gas \citep[see e.g. fig. 8 of][]{Crain2015}. Due to the high $\A$ sequence of interest being concentrated towards the centre of these galaxies, we consider just the bulge of the galaxy in Section \ref{5}.

\section{The fraction of field stars in the bulge originating in GCs\label{5}}

\subsection{Bulge stars from disrupted GCs in E-MOSAICS}

\begin{figure}
	\includegraphics[width=\linewidth]{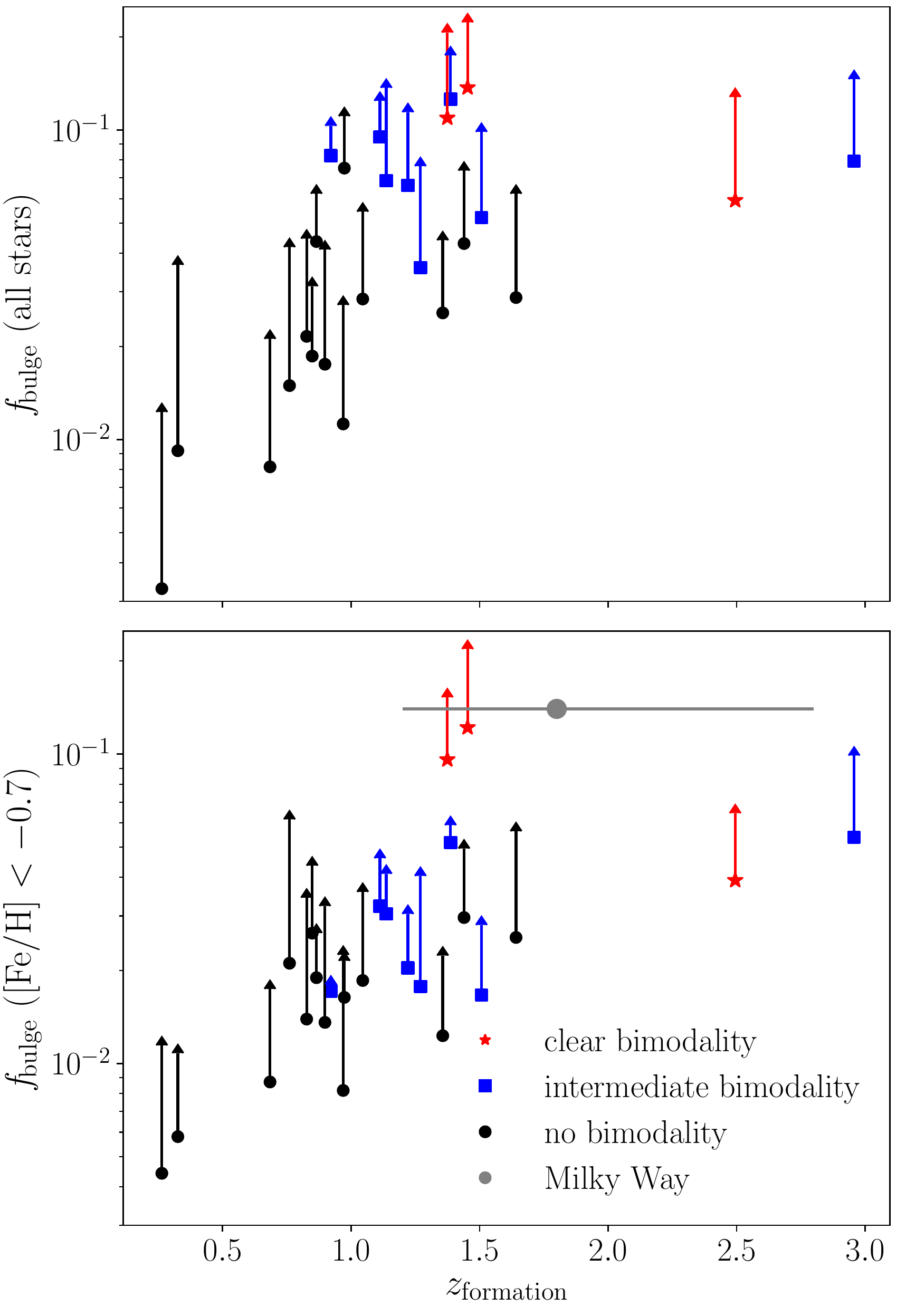}
    \caption{The fraction of field stars in the bulge which were born in GCs ($f_\mathrm{bulge}$) as a function of the redshift ($z_{\rm formation}$) at which all progenitors together have attained half of the $z = 0$ stellar mass (this is effectively the median age or median formation redshift of all stars in the galaxy at $z = 0$). The different symbols represent the degree of bimodality the field stars show in the $\Oxy$ - $\FeH$ plane, from definite bimodality (stars), to intermediate bimodality (squares) to no bimodality at all (circles). The data are shown as lower limits due the under disruption of GCs in E-MOSAICS. The upper limits (at the tips of the arrows) show the extreme assumption that no GC survives in the bulge. Top panel: $f_\mathrm{bulge}$ across all field stars in the bulge. Bottom panel: $f_\mathrm{bulge}$ for field stars below $\FeH < -0.7$ to match the selection when deriving $f_\mathrm{bulge}$ for the MW. The grey data point indicates the bulge mass fraction from GCs derived by \citet{Schaivon2017} for the formation redshift of the Milky Way inferred from the age-metallicity distribution of its GC population by \citet{kruijssen19b}.}
    \label{fig:fig7.pdf}
\end{figure}

We found in Section \ref{4} that the high $\A$ sequence of stellar particles shown in Fig. \ref{fig:big_plot.pdf} mostly reside close the centre of the galaxies. We therefore target the bulges of the 25 E-MOSAICS galaxies in the rest of this analysis, with a focus on the contribution of disrupted GCs to the formation of the bulge.  Since the size of the bulge varies for each galaxy we make a radius and orbital circularity cut. We define the orbital circularity as in \cite{Abadi2003}, $\epsilon_\mathrm{J} = J_\mathrm{z}/J_\mathrm{c} (E)$ (i.e. the angular momentum relative to the angular momentum of a circular orbit), where $\epsilon = 1$ describes a perfectly circular orbit. The field stars in the bulge are therefore defined as the stellar particles within the stellar half mass radius with $\epsilon_{\mathrm{J}}< 0.5$ \citep{Sales2015}.\par

We can now determine the contribution of disrupted GCs to the stellar population of the bulge, in the form of the fraction of field stars in the bulge that were born in a GC ($f_\mathrm{bulge}$). The E-MOSAICS galaxies host too many GCs at $z=0$ due to under disruption \citep{Pfeffer2018,kruijssen19}. For this reason, our fractions should be considered as lower limits. Therefore we also show the extreme upper limits on $f_\mathrm{bulge}$, where we assume that every GC formed in the bulge of the galaxy becomes disrupted i.e. no GC survives to the present day in the bulge. Such an extreme assumption does not affect the general trend of the simulations. \par

We show $f_\mathrm{bulge}$ for all 25 galaxies in the top panel of Fig. \ref{fig:fig7.pdf} and see that there is a large range in the value of $f_\mathrm{bulge}$, from 0.3-14 per cent. This fraction is dependent on the time at which the total stellar mass of all progenitors of that galaxy reaches half of the $z=0$ mass ($z_\mathrm{formation}$, \citealt{DeLucia2006,Qu2017}), such that galaxies that formed faster have higher $f_\mathrm{bulge}$.\par

We also group the galaxies by the shape of their stellar contours in $\A$-$\FeH$ space (like those shown in Fig. \ref{fig:row.pdf}). We place each galaxy into one of three categories:
\begin{itemize}
\item No bimodality, where there is a smooth decline in $\Oxy$ for increasing $\FeH$ (e.g. MW06).
\item Intermediate bimodality, where there is a small bump or a slight increase in $\Oxy$ for a given metallicity (e.g. MW23 and MW12).
\item Clear bimodality, where there are two distinct $\Oxy$ sequences at a given metallicity (e.g. MW13 and MW18).
\end{itemize}
We conclude that the most bimodal galaxies are those that have faster formation times and higher $f_\mathrm{bulge}$. This is to be expected since, from Section \ref{4}, the high $\A$ sequence in the bimodal galaxies is formed from a high pressure environment close to the centre of the galaxy where GCs are efficiently formed and subsequently disrupted.
\par

\begin{figure}
	\includegraphics[width=\linewidth]{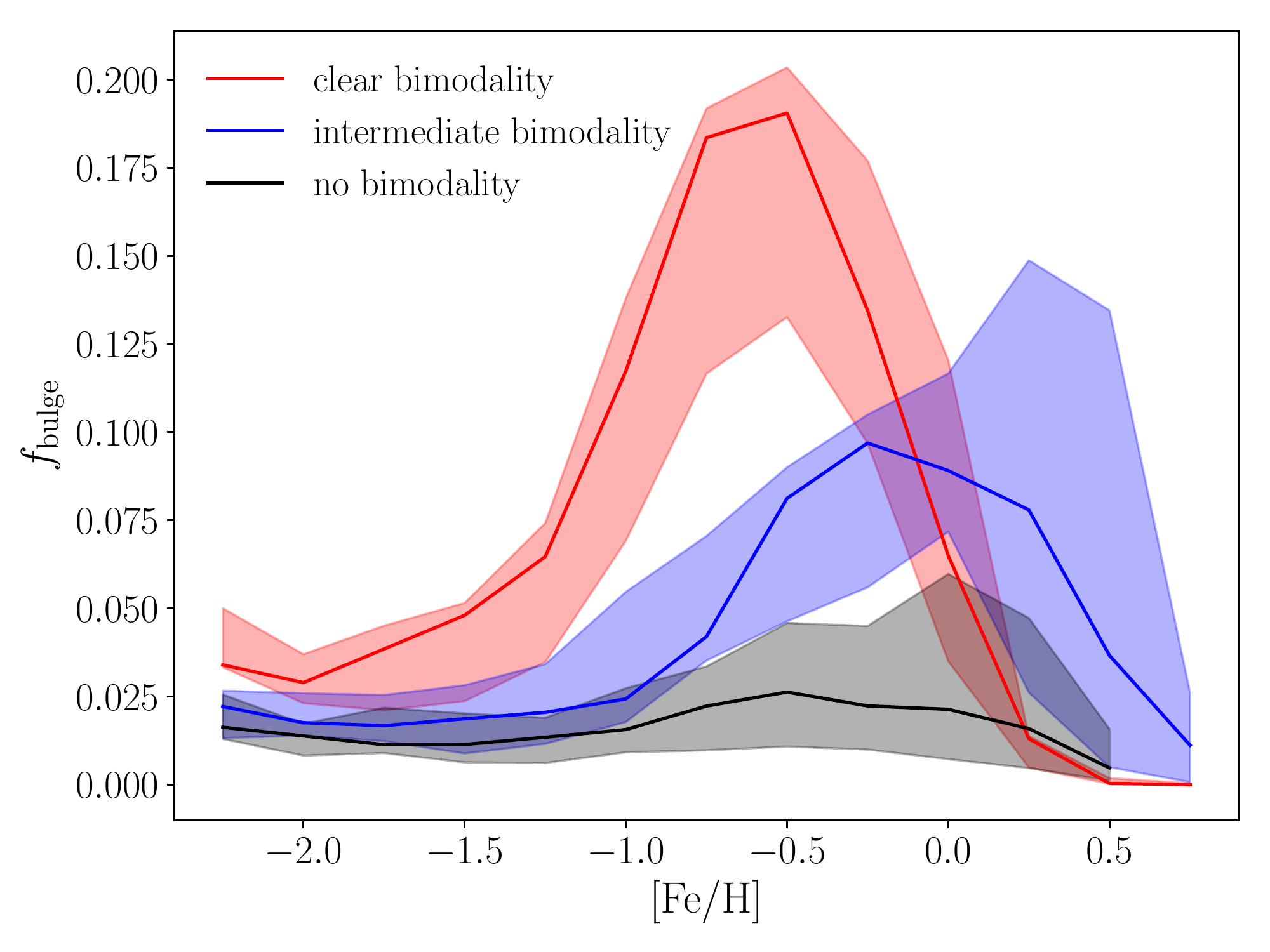}
    \caption{The fraction of field stars in the bulge which were contributed by GCs ($f_\mathrm{bulge}$) in 0.5 dex $\FeH$ bins. The solid lines show the running median and the shaded regions represent the 16th-84th percentile range. The galaxies are stacked by the level of bimodality their stars show in the $\A$-$\FeH$ plane, coloured as in Fig. \ref{fig:fig7.pdf}. $f_\mathrm{bulge}$ peaks higher and at lower metallicities in galaxies with clear $\A$ bimodality. }
    \label{fig:met_bins.pdf}
\end{figure}

We also present the metallicity dependent $f_\mathrm{bulge}$  in Fig. \ref{fig:met_bins.pdf}. Here, $f_\mathrm{bulge}$ is calculated in $\FeH$ bins of 0.5 dex and the median and the 16th-84th percentile range of the galaxies grouped by their level of bimodality is shown. For all galaxies, GC disruption and the contribution of GCs to the field star population increases towards higher metallicites for $\FeH < -0.5$. The clearly bimodal galaxies show the highest and most metal-poor peak. The intermediate bimodal galaxies show a smaller and slightly more metal-rich peak and the galaxies with no bimodality show a relatively flat $f_\mathrm{bulge}$ with increasing $\FeH$.  Fig. \ref{fig:fig7.pdf} allows us to conclude that the MW has a high fraction of disrupted GCs in its bulge, consistent with the most bimodal simulated galaxies. We therefore use the most bimodal galaxies in Fig. \ref{fig:met_bins.pdf} to predict that the stars from disrupted GCs in the bulge of the MW will show the highest fraction around $\FeH \approx -1$.  

\subsection{Comparison with the Milky Way}
Searching for stars from disrupted GCs in the bulge of the MW is something that has been done with large scale surveys of galactic stars. These searches mainly focus on finding populations of nitrogen rich stars \citep{Schaivon2017}. 

Whilst some stars in GCs show the same chemical abundances as those found in the field (first population, FP), others show enhancements or depletions in some elements, such as a nitrogen enhancement (second population, SP, e.g. \citealt{Carretta2009}, for a recent review see \citealt{Bastian2018}).\par

It is interesting here to address the effect the formation time-scales of multiple populations may have on our comparison with the MW bulge. The main theories for the origin of multiple populations in GCs suggest that the formation of SP stars happens on timescales $\mathrm{<300 \ Myr}$. For scenarios invoking enrichment by massive stars, the timescales for SP star formation are $\mathrm{< 5-10 \ Myr}$ (e.g. \citealt{Gieles2018}). For the asymptotic giant branch (AGB) scenario, the timescale to form SP stars is $\mathrm{< 100 \ Myr}$, as this is the timescale for the first Type Ia SNe to stop further generations of stars forming within the cluster \citep{D'Ercole2008}. Furthermore, observations also give us some constraint on the relative formation times of FP and SP stars. Ancient GCs show that the age difference between the two populations is $\mathrm{< 200 \ Myr}$ and consistent with $\mathrm{0 \ Myr}$ \citep{Marino2012}. Younger GCs also show a similar difference with an age difference calculated for NGC 1978 of $\mathrm{1 \pm 20 \ Myr}$ \citep{Martocchia2018b}. Therefore, both theoretical and observational evidence suggest that GC disruption happens after multiple populations form and we can therefore directly compare the nitrogen rich stars in the MW with disrupted GCs in the simulations without adding any extra calculations for the formation times of the nitrogen rich stars.\par

\cite{Schaivon2017} used APOGEE to find a population of nitrogen rich stars in the bulge of the MW. Restricting their selection to stars below $\FeH = -1$, they found that 7 percent of the field stars in the bulge are nitrogen enhanced. Although they advance multiple explanations for the origin of these stars, their preferred explanation is that they are remnants of disrupted GCs. \cite{Schaivon2017} then assume equal numbers of first and second population stars are lost from GCs and calculate that 14 percent of field stars in the bulge (within this metallicity selction) formed in a GC.

In order to fairly compare the results from the E-MOSAICS simulation to that of the MW, we also make a metallicity selection for the stars in the bulges of the simulations. Due to the metallicities in the EAGLE simulations being overestimated \citep{Schaye2015} we make a metallicity cut of $\FeH < -0.7$ to compare with the metallicity cut of $\FeH < -1$ in \citet{Schaivon2017}. We show this in the bottom panel of Fig. \ref{fig:fig7.pdf}, using for the MW, $z_\mathrm{formation} = 1.8^{+1.0}_{-0.6}$ \citep{kruijssen19b}. Although the metallicity cut reduces the $f_\mathrm{bulge}$ calculation for most galaxies, the MW is consistent with the subset of E-MOSAICS galaxies with a bimodal $\A$ distribution, fitting with the observation that the MW also exhibits an $\A$ bimodality \citep{Fuhrmann1998,Adibekyan2012,Hayden2015}.  We note here that \citet{Schaivon2017} did not make any orbital cuts in their selection, however when we do not include the circularity cut, there is no significant change in Fig. \ref{fig:fig7.pdf} and therefore we keep this in our analysis for consistency.\par

\begin{figure}
	\includegraphics[width=\linewidth]{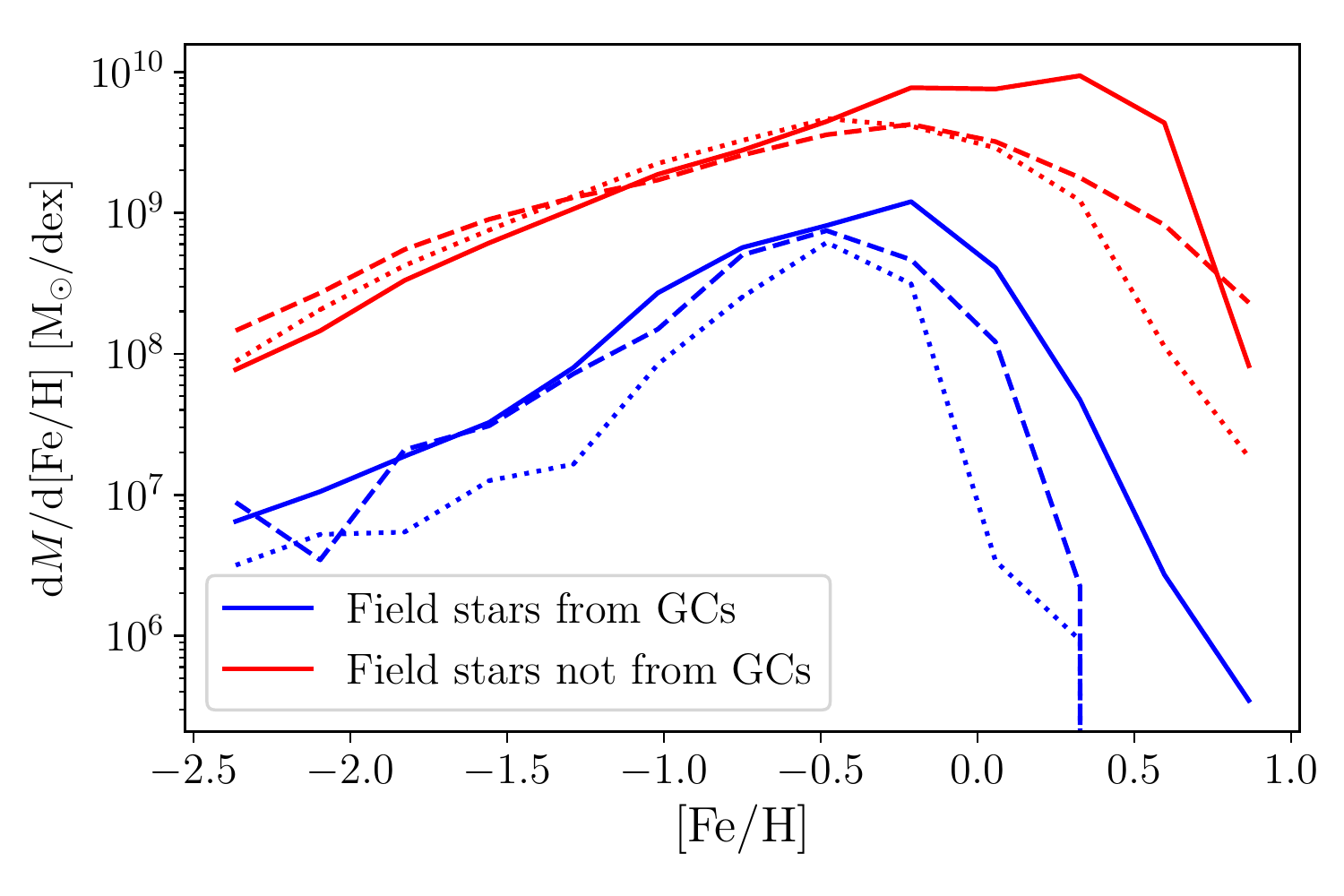}
    \caption{The metallicity distribution function of field stars in the bulges of the three E-MOSAICS galaxies that show bimodal $\A$ at a given $\FeH$, these are MW13 (solid line), MW18 (dashed line) and MW21 (dotted line). The stars are split by whether they were born in a GC (blue lines) or not (red lines).  }
    \label{fig:metallicity_bimodalGal.pdf}
\end{figure}

\cite{Schaivon2017} also present the metallicity distribution of their sample of nitrogen enriched stars. The metallicity distribution function (MDF) peaks at $\FeH \sim -1$ whereas the MDF of the bulge field stars begins to decline at this metallicity (\citealt{Schaivon2017}, fig. 9). Therefore we conclude that for the MW $f_\mathrm{bulge}$ will indeed peak at $\FeH \sim -1$ as predicted from the simulations in the previous section. In Fig. \ref{fig:metallicity_bimodalGal.pdf},  we present the metallicity distribution function of field stars in the bulge of the three galaxies in our sample that are classified as bimodal in their $\A$ distributions (MW13, MW18 and MW21). The field stars are coloured by whether they were once bound to GC (blue lines) or not (red lines). \citet{Schaivon2017} present a similar figure in which they show the metallicity distributions of nitrogen-rich and nitrogen-normal stars. Similar to the distribution shown in \citet{Schaivon2017} the ex-GC stars in the E-MOSAICS simulations show a more peaked distribution than that of the field stars not from GCs. Also similar to \citet{Schaivon2017} the ex-GC stars peak at a lower metallicity than the stars not born in GCs (albeit at a higher metallicity than the MW). \par

Our findings therefore corroborate the conclusion of \cite{Schaivon2017} that the population of nitrogen enriched stars found in the bulge of the Milky Way are likely to be from disrupted clusters. However, such a large fraction (14 percent) of stars in the bulge originating in clusters is rare in the E-MOSAICS simulations, with only 2 of the galaxies showing fractions greater than 10 per cent. In the context of our findings, such a high nitrogen-enhanced fraction of bulge stars constitutes further evidence that the MW formed unusually early in cosmic history, given its halo mass (also see \citealt{Mackereth2018} and \citealt{kruijssen19b}).\par

\section{Conclusions\label{6}}
This work uses the E-MOSAICS simulations to investigate the $\A$-$\FeH$ distribution of a galaxy's field stars and GCs. Fig. \ref{fig:row.pdf} reveals many interesting features in the $\A$-$\FeH$ space of the E-MOSAICS galaxies, namely, bimodal distributions and a lack of GCs in the high $\alpha$ sequence. Therefore, we use 25 MW-like galaxies and their GC populations to understand what their $\A$-$\FeH$ distribution can reveal about the formation of the galaxy, both in terms of the $\alpha$ abundance ratios of individual GCs and where in $\A$-$\FeH$ space we should look for remnants of disrupted GCs.  \par

Many works present the hypothesis that GCs should follow the $\A$-$\FeH$ distribution of field stars and those GCs which have been accreted should show relatively low $\A$ for their $\FeH$.  We show in Fig. \ref{fig:insitu_exsitu.pdf} that the GCs do follow the general trend of the field stars and if we were to observe a GC with a low $\A$ abundance, then it is likely that this GC has been recently accreted alongside a dwarf galaxy. However, it is impossible to distinguish between in-situ GCs and GCs that were accreted early in the formation history of the galaxy based on $\alpha$ enhancement alone.\par

When focusing on the field star $\A$-$\FeH$ distribution, there is a wide range of shapes, from a smooth decline to clearly bimodal (Fig. \ref{fig:row.pdf}). The high-$\alpha$ field star sequence present in some of our galaxies is made up of a large fraction of disrupted GCs (Fig. \ref{fig:big_plot.pdf}). This is due to the high pressure environment that is necessary to create a high $\A$ sequence. This environment creates very short gas consumption times ($T_\mathrm{g}$), making it ideal for GC formation and, due to the `cruel cradle effect', subsequent destruction. This area of high $\A$ is located close to the centre of the galaxy and we therefore calculate the fraction of disrupted GCs contributing to the bulge of each of the 25 galaxies ($f_\mathrm{bulge}$). \par

Fig. \ref{fig:fig7.pdf} shows that the galaxies which show the strongest bimodality also show rapid early growth of their progenitors. It is also the most bimodal galaxies that have the highest contribution from disrupted clusters, $f_\mathrm{bulge}$. \citet{Mackereth2018} showed that a high-$\alpha$ sequence in MW-like galaxies is formed via a phase of rapid early formation, a conclusion that is corroborated by our comparison of the $f_\mathrm{bulge}$-$z_\mathrm{formation}$ relation with the high $f_\mathrm{bulge}$ fraction inferred by \cite{Schaivon2017}. We also add that galaxies that formed, on average, earlier than typical galaxies of that mass, are likely to have a relatively high fraction of stars in the bulge that originated in GCs. It is the high-$\alpha$ sequence in these galaxies are likely to contain a high fraction of stars that were born in GCs.

We compare the $f_\mathrm{bulge}$ of the E-MOSAICS galaxies to that of the Milky Way and find that the Milky Way has an unusually high $f_\mathrm{bulge}$, comparable to only 2 out of the 25 E-MOSAICS galaxies. This is consistent with the conclusions of previous works that the Milky Way underwent a period of rapid growth early in its formation, suggesting that its mass assembly history is atypical of $L_\star$ galaxies (e.g. \citealt{Mackereth2018,kruijssen19b}).

\section*{Acknowledgements}
We thank the anonymous reviewer for a positive report that improved the clarity of this paper. JP and NB gratefully acknowledge funding from a European Research Council consolidator grant (ERC-CoG-646928-Multi-Pop). JMDK gratefully acknowledges funding from the German Research Foundation (DFG) in the form of an Emmy Noether Research Group (grant number KR4801/1-1). JMDK and MRC gratefully acknowledge funding from the European Research Council (ERC) under the European Unions Horizon 2020 research and innovation programme via the ERC Starting Grant MUSTANG (grant number 714907). MRC is supported by a Fellowship from the International Max Planck Research School for Astronomy and Cosmic Physics at the University of Heidelberg (IMPRS-HD). NB and RAC are Royal Society University Research Fellows. This work used the DiRAC Data Centric system at Durham University, operated by the Institute for Computational Cosmology on behalf of the STFC DiRAC HPC Facility (www.dirac.ac.uk). This equipment was funded by BIS National E-infrastructure capital grant ST/K00042X/1, STFC capital grants ST/H008519/1 and ST/K00087X/1, STFC DiRAC Operations grant ST/K003267/1 and Durham University. DiRAC is part of the National E-Infrastructure. The study also made use of high performance computing facilities at Liverpool John Moores University, partly funded by the Royal Society and LJMU's Faculty of Engineering and Technology.




\bibliographystyle{mnras}
\bibliography{bibliography.bib} 

\appendix
\section{Link between bimodality and gas consumption time\label{A}}
\begin{figure*}
	\includegraphics[width=\linewidth]{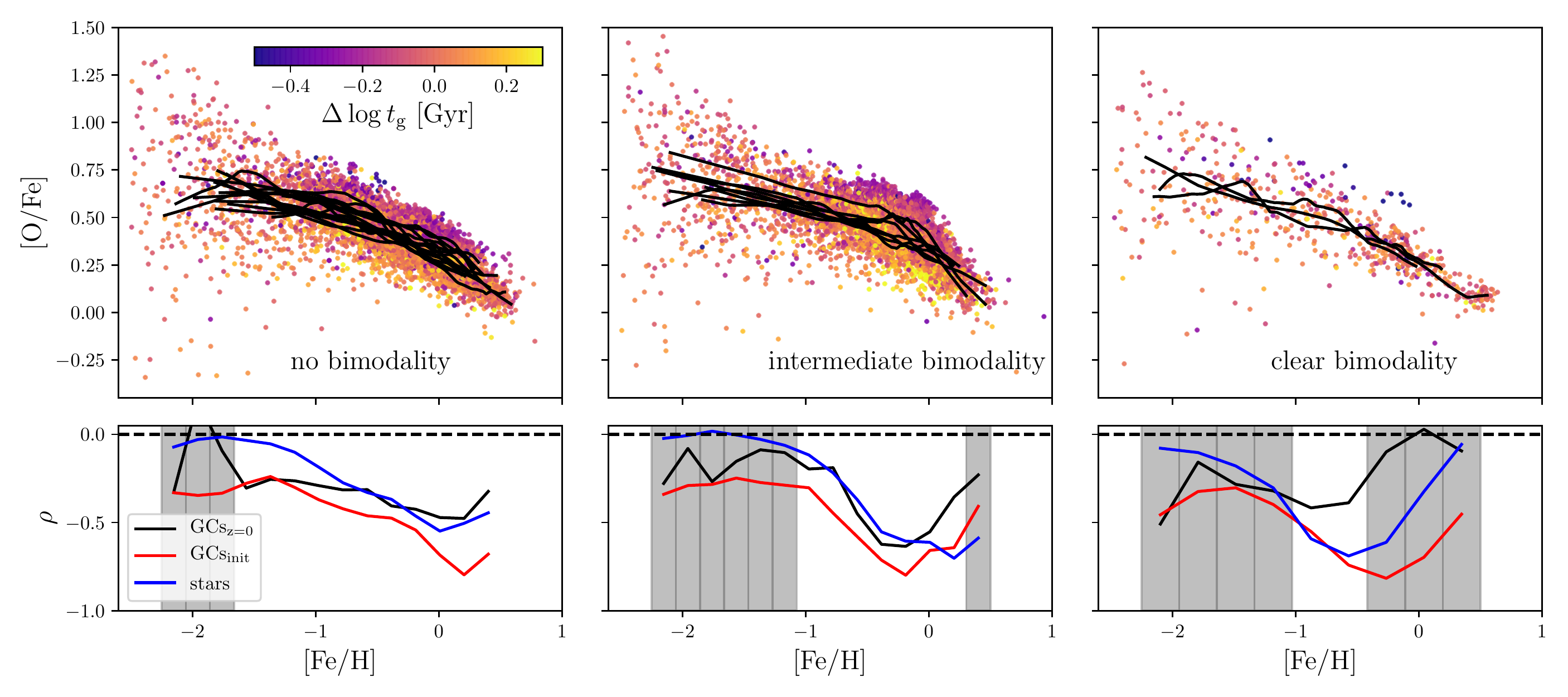}
    \caption{Top row: the $\Oxy$-$\FeH$ relation for GCs grouped by their galaxy's degree of bimodality. The solid black lines represent the running medians of the individual galaxies, computed using the LOWESS method. Each point represents one $z=0$ GC coloured by its difference from the running median of the $t_\mathrm{g}$-$\FeH$ relation of its host galaxy. Bottom row: the Spearman rank correlation coefficent of the $\Delta t_\mathrm{g}$-$\Delta \Oxy$ relation. Each line represents present day GCs ($\mathrm{GCs_{z=0}}$, black), any GC that formed ($\mathrm{GCs_{init}}$, red) and the field stars (blue). The shaded regions highlight the metallicity bins in which the Spearman correlation coefficient for $\mathrm{GCs_{z=0}}$ is not significant (Spearman p-value > 0.01). The other populations have significant correlations in all $\FeH$ bins. The $\FeH$ bins are wider in the clearly bimodal case to account for lower number statistics.}
    \label{fig:tg_all.pdf}
\end{figure*}

The high $\Oxy$ sequence observed in some of the E-MOSAICS galaxies can be explained via a high pressure natal environment. In a high density (pressure) environment, the gas consumption timescale of the natal gas is short, and therefore the natal gas is consumed before it can be enriched with the $\mathrm{Fe}$ nucleosynthesised by Type Ia supernovae. \citet{Mackereth2018} shows that the amount of $\alpha$ enhancement correlates with the gas consumption timescale ($t_\mathrm{g}$) of field stars. We now test this for the GCs and field stars in the 25 E-MOSAICS galaxies. The consumption time of the natal gas from which the GC formed is calculated, following the description in \citet{Schaye2008} (eq. 11), as
\begin{equation}\label{eq:3}
t_\mathrm{g} = A^{-1}(1 M_\mathrm{\odot} \mathrm{pc}^{-2})^{n}\Big(\frac{\gamma}{G} f_{g} P_{\star}\Big)^{(1-n)/2}.
\end{equation}
Here, the parameters $A=1.515 \times 10^{-4} M_{\odot} yr^{-1} kpc^{-2}$ and $n=1.4$ are specified by observations, see \citet{Schaye2015} for details. $\gamma = 5/3$ is the ratio of specific heats for an ideal gas, $f_g$ is the local gas fraction (assumed to be unity) and $P_\star$ is the pressure of the natal gas. $t_\mathrm{g}$ provides an estimate of the amount of time a star forming gas particle resides in the ISM before becoming a star particle (though we note this is an over-estimate since it neglects ejection of the ISM in winds), a low $t_\mathrm{g}$ therefore indicates vigorous star formation.\par

Following the methodology used by \citet{Davies2019}, for each of the galaxies we compute the running median of the $\Oxy$-$\FeH$ relation and the $t_\mathrm{g}$-$\FeH$ relation using the locally weighted scatter plot smoothing method (LOWESS, \citealt{Cleveland1979}). The running medians are calculated separately for the stars, GCs and initial GCs. We then compute the difference from these running medians for each GC and star, i.e. $\Delta t_\mathrm{g, GC} = t_\mathrm{g, GC} - t_\mathrm{g, GC, median}$. The correlation between $\Delta \Oxy$ and $\Delta t_\mathrm{g}$ is computed as a Spearman rank correlation coefficient ($\rho$).\par

The top row of Fig. \ref{fig:tg_all.pdf} shows the $z=0$ GCs along with the running medians (black lines) for each of the galaxies. Each GC is coloured by the consumption time of its natal gas ($t_\mathrm{g}$). The bottom panels of Fig. \ref{fig:tg_all.pdf} show the Spearman-$\rho$ values for the $z=0$ GCs shown in the top panels and also the initial GCs and the field stars. Here the z=0 GCs are those that survive until present day with a mass $> 10^5 \mathrm{M_{\odot}}$, the initial GCs are those that were formed with a mass $> 10^5 \mathrm{M_{\odot}}$ but do not necessarily survive until $z=0$. \par

In the left and the middle panels of Fig. \ref{fig:tg_all.pdf} the $\rho$-values are calculated in 15 equally sized bins from $-2.5 < \FeH < 0.5$. In the right panel the $\rho$-values are calculated in 10 equally sized bins across the same metallicity range, to account for lower number statistics. 
The Spearman p-value is indicative as to whether a correlation is significant as it depends on the strength of the correlation and the sample size. We calculate the p-value for each of the bins and shade the bins to highlight where the $\Delta \Oxy$-$\Delta t_\mathrm{g}$ correlation is not significant (the Spearman-p value exceeds 0.01). For the $z=0$ GCs, the correlation for the initial GCs and the stars is significant everywhere. The galaxies are grouped by the shape of their field star contours in $\Oxy$-$\FeH$ space (as discussed in Section \ref{5}) and from left to right show: no bimodality, intermediate bimodality and clear bimodality.\par

Inspection by eye shows that, in the top panels, the GCs that have a higher than average $\Oxy$ have a lower than average $t_\mathrm{g}$. Hence, the recovered relation is negatively correlated ($\rho < 0$) for much of the range in $\FeH$. Section \ref{4} discusses how the high $\Oxy$ field star sequence in the bimodal galaxies is formed in high pressure environments that induce a short gas consumption time. This is seen directly in Fig. \ref{fig:tg_all.pdf}, the field stars show a negative correlation everywhere above an $\FeH > -1$. However, the correlation is stronger in the intermediate and clearly bimodal galaxies. The initial GCs show the strongest correlation. They show a stronger correlation than the field stars because some stars that form with fast consumption times are $\alpha$-poor, this therefore weakens the field star correlation.\par

As discussed in Section \ref{4}, the high pressures that create the most $\alpha$-rich stars in the clearly bimodal galaxies creates the perfect environment to form and then subsequently destroy high $\alpha$ GCs. Again this can be seen directly in Fig. \ref{fig:tg_all.pdf}. In the lower right panel the initial GCs show a strong negative correlation between their $\alpha$ enhancement and relative $t_\mathrm{g}$. However, the GCs that survive until the present day do not show such a relation, because many of the GCs with the shortest $t_\mathrm{g}$ have been destroyed. All galaxies show a weaker correlation between $\Delta \Oxy$ and $\Delta t_\mathrm{g}$ in their $z=0$ GCs when compared to their initial GCs, but the difference is most pronounced in the most bimodal galaxies. At $\FeH = -0.5$ (where the correlation is significant in all galaxies) we calculate the difference in the Spearman-$\rho$ value between the initial GCs and final GCs for the 'clearly bimodal' and 'no bimodality' galaxies. The clearly bimodal galaxies have a $\rho$-value difference of 0.44 whereas the galaxies with no bimodality have a difference of 0.11.

\bsp	
\label{lastpage}
\end{document}